\documentclass{article}
\usepackage{graphicx}
\usepackage{amsmath}

\begin{document}

\title{Second Comment on ``Contextuality within quantum mechanics manifested in
subensemble mean values'' [Phys. Lett. A 373 (2009) 3430]}
\author{F. De Zela \\
Departamento de Ciencias, Secci\'{o}n F\'{i}sica \\
Pontificia Universidad Cat\'{o}lica del Per\'{u}, Av. Universitaria 1801,
Lima 32, Peru.}
\maketitle

\begin{abstract}
I examine Pan and Home's reply to my Comment on their proposal for
testing noncontextual models. I show that the Kochen-Specker model
for a qubit does explain all outcomes of a test based on such a
proposal, so that it would be inconclusive about the untenability
of realistic, noncontextual models.

\end{abstract}

Pan and Home (PH) introduced recently \cite{pan} what they claimed
to be a new type of contextuality between the spin and path
degrees of freedom of a particle. Their goal was to show that the
formalism of Quantum Mechanics (QM) embodies this kind of
contextuality, in the sense that subensemble mean values of spin
measurements are ``contingent upon what \emph{choice} is made of
measuring a suitably defined comeasurable (commuting) `path'
observable'' \cite{pan}. PH discussed a variant of a Mach-Zehnder
interferometer, by means of which the alleged contextuality could
be exhibited. In a Comment to PH's paper \cite{fdz} I tried to
show that such a setup would be essentially equivalent to a
standard Stern-Gerlach (SG) array, in the sense that any
experiment performed with PH's setup could be replicated with a SG
array. As all outcomes of the latter could be explained with the
Kochen-Specker model for a qubit, so would also be the case for a
test based on PH's proposal. In a reply to my Comment Pan and Home
argue \cite{pan2} that this is not so, because their arrangement
allows path-measurements that the SG setup does not. Such
path-measurements could be effectively performed in PH's array by
recording which one of two SG devices has detected a spin-1/2
particle. Now, Pan and Home reproduce in their Reply \cite{pan2}
the setup I proposed; but without including a SG device which
serves to prepare the spin-states I have considered. It is this
missing SG device the one which carries the information that PH
ascribe to their adjustable beam splitter. Last one allows to fix
the different path degrees of freedom. Hence, the apparently
missing stage in the array I considered, and that could serve to
effectuate a ``path-measurement'', was actually there. Indeed, the
SG device at the entrance of this array does select one of two
paths. The selected path carries then the same information as in
PH's array. This is reflected in the subensemble mean values of
spin measurements that can be recorded at the two arrays -- PH's
and mine -- which are the same.

My aim has been to show that PH's setup is physically equivalent
to a SG setup for the sake of testing realistic models. As a
consequence, the Kochen-Specker (KS) model for a qubit
\cite{kochen} could be invoked to explain all possible outcomes of
an experiment implementing PH's proposal. Now, it is certainly
unnecessary to make the detour of showing that one setup is
physically equivalent to the other, if our aim is simply to show
that the KS model does explain all outcomes of some given setup.
The aim of this note is to directly show how the KS model can
explain the outcomes of an experiment performed according to PH's
proposal.

Let us then assume that an experimental group has implemented PH's
proposal, obtaining results that are in full agreement with the
corresponding quantum-mechanical predictions. The purpose of
such an experiment is to rule out some class of realistic
theories. In the present case this class is constituted by
noncontextual realistic theories, or models. In other words, no
such a model should serve to explain the experimental outcomes.
But, as we shall see, there is a model, the KS model, that would
be capable of explaining all these outcomes. This is so because
the KS model is capable of reproducing all quantum-mechanical
predictions concerning a two-state Hilbert space, which is the
only one involved in the \emph{measurements} of PH's
proposal. Let us remind that observables in such a space are of the form $\widehat{A}=a_{0}I+%
\overrightarrow{a}\cdot \overrightarrow{\sigma }$, where $\overrightarrow{%
\sigma }$ represents the triple of Pauli matrices and $I$ the
identity matrix.

Briefly, the experiment that realizes PH's proposal consists in
submitting to the action of a Mach-Zehnder-like array a beam of
neutrons whose spins are polarized along the $+\widehat{z}$-axis,
i.e., neutrons being prepared in the spin-up state $\left|
\uparrow \right\rangle _{z}$. The Mach-Zehnder-like array consists
of two beam splitters, some mirrors, a
spin-flipper, two SG devices and four detectors (see Fig.1 in Ref.\cite{pan}%
). The experimental outcomes refer to subensemble spin mean values $%
\left\langle \widehat{\sigma }_{\theta }\right\rangle _{SG}$ that
are drawn from the detectors set at the output of the SG devices.
These SG devices can be freely oriented, $\theta $ being an angle
fixing the orientation. One of the two beam splitters in the
Mach-Zehnder-like array is a $50:50$ beam splitter while the other
is one of adjustable reflectivity/transmissivity, whose action can
be represented by

\begin{equation}
A_{\gamma }=\left(
\begin{array}{cc}
\left( \gamma ^{2}-\delta ^{2}\right)  & -2i\gamma \delta  \\
2i\gamma \delta  & -\left( \gamma ^{2}-\delta ^{2}\right)
\end{array}
\right) ,  \label{1}
\end{equation}
with $\gamma $ and $\delta $ being reflection and transmission
(real) coefficients satisfying $\gamma ^{2}+\delta ^{2}=1$. The
matrix representation of $A_{\gamma }$ refers to a path-space
basis $\left\{ \left| \psi _{1}\right\rangle ,\left| \psi
_{2}\right\rangle \right\} $ constituted by the path states
associated to the two input ports of the adjustable beam splitter.
Its output states are $\left| \psi _{3}\right\rangle =-i\gamma
\left| \psi _{1}\right\rangle +\delta \left| \psi _{2}\right\rangle $ and $%
\left| \psi _{4}\right\rangle =\delta \left| \psi _{1}\right\rangle -i\gamma
\left| \psi _{2}\right\rangle $, in terms of which $A_{\gamma }$ is defined
as $A_{\gamma }\equiv \left| \psi _{3}\right\rangle \left\langle \psi
_{3}\right| -\left| \psi _{4}\right\rangle \left\langle \psi _{4}\right| $.

The quantum-mechanical predictions that the experiment should confirm are
given by

\begin{eqnarray}
\left\langle \widehat{\sigma }_{\theta }\right\rangle _{SG1} &=&\frac{1}{2}{%
\vphantom{\left\langle \downarrow \right|}}_{\vartheta }{\left\langle
\downarrow \right| }\widehat{\sigma }_{\theta }\left| \downarrow
\right\rangle _{\vartheta }=-\frac{1}{2}\cos \left( 2\left( \vartheta
-\theta \right) \right) ,  \label{sg1} \\
\left\langle \widehat{\sigma }_{\theta }\right\rangle _{SG2} &=&\frac{1}{2}{%
\vphantom{\left\langle \uparrow \right|}}_{\vartheta }{\left\langle \uparrow
\right| }\widehat{\sigma }_{\theta }\left| \uparrow \right\rangle
_{\vartheta }=+\frac{1}{2}\cos \left( 2\left( \vartheta -\theta \right)
\right) .  \label{sg2}
\end{eqnarray}
Here, I have set $\gamma =\sin \vartheta $ and $\delta =\cos
\vartheta $. The mean values refer to states $\left| \uparrow
\right\rangle _{\vartheta }=\sin \vartheta \left| \downarrow
\right\rangle _{z}+\cos \vartheta \left| \uparrow \right\rangle
_{z}$ and $\left| \downarrow \right\rangle _{\vartheta }=\cos
\vartheta \left| \downarrow \right\rangle _{z}-\sin \vartheta
\left| \uparrow \right\rangle _{z}$. These states are prepared by
choosing an appropriate value of $\vartheta $, viz., of $\gamma $ (or $%
\delta $).

Let us now turn to a KS model that explains the above results.
Being a realistic model,
the KS model assumes that any quantum-mechanical state $%
\left| \psi \right\rangle $ conveys incomplete information about a
physical system, and this should be the reason why we cannot
predict with certainty the results of measurements performed on
the system. This means that $\left| \psi \right\rangle $
represents in fact a whole family of systems, whose members could
be in principle distinguished from one another by a series of
supplementary parameters $\lambda $, so-called ``hidden
variables''. In the case of a two-level system we can generally
write $\left| \psi \right\rangle =\cos \left( \theta _{\psi
}/2\right) $ $e^{-i\varphi _{\psi }/2}\left| +\right\rangle +\sin
\left( \theta _{\psi }/2\right) $ $e^{i\varphi _{\psi }/2}\left|
-\right\rangle $, with $\left| \pm \right\rangle $ being the
eigenvectors of $\sigma _{z}$. Within the formalism of QM the
states $\left| \psi \right\rangle $ and $e^{i\alpha }\left| \psi
\right\rangle $ represent one and the same physical state. This
state is thus more properly represented by an equivalence class,
$\{\left| \psi ^{\prime }\right\rangle =e^{i\alpha }\left| \psi
\right\rangle \sim \left| \psi \right\rangle \}$, or by the
projector $\left| \psi \right\rangle \left\langle \psi \right|
=\left( I+\widehat{n}_{\psi }\cdot \overrightarrow{\sigma }\right) /2$, with $%
\widehat{n}_{\psi }=(\sin \theta _{\psi }\cos \varphi _{\psi },\sin \theta
_{\psi }\sin \varphi _{\psi },\cos \theta _{\psi })\in S^{2}$. The KS model
takes as hidden variables the vectors $\widehat{n}_{\lambda }=(\sin \theta
_{\lambda }\cos \varphi _{\lambda },\sin \theta _{\lambda }\sin \varphi
_{\lambda },\cos \theta _{\lambda })$ on the unit sphere $S^{2}$. Given the
state $\left| \psi _{a}\right\rangle \left\langle \psi _{a}\right| =\left( I+%
\widehat{n}_{a}\cdot \overrightarrow{\sigma }\right) /2$, say, the KS model
assigns to it the following probability for its occurrence:

\begin{equation}
\rho _{a}(\lambda )d\lambda =\frac{\widehat{n}_{\lambda }\cdot \widehat{n}%
_{a}}{\pi }\Theta \left( \widehat{n}_{\lambda }\cdot \widehat{n}_{a}\right)
d\lambda ,  \label{ro}
\end{equation}
with $\Theta $ meaning the Heaviside's step-function and $d\lambda
$ the uniform measure on the sphere ($d\lambda =\sin \theta
_{\lambda }d\theta _{\lambda }d\varphi _{\lambda }$). It can then
be proved \cite{fdz} that for any $\widehat{A}=a_{0}I+%
\overrightarrow{a}\cdot \overrightarrow{\sigma }$ and $\left| \psi
\right\rangle $ there is a function $A(\lambda )$ such
that$\left\langle \psi \right| \widehat{A}\left| \psi
\right\rangle =\int \rho _{\psi }(\lambda )A(\lambda )d\lambda $.

Focusing on PH's proposal, a KS model tailored to explain its
outcomes can be as follows. Neutrons entering PH's array are
described by a probability distribution like that of Eq.(\ref{ro})
with $\widehat{n}_{a}=\widehat{z}$. The effect that
the Mach-Zehnder part of PH's array has on neutrons -- so the model's prescription -- is to flip $%
\widehat{z}$. Those neutrons exiting the array through port 3 of the
adjustable beam-splitter (see Fig.1 in Ref. \cite{pan}) are in a state whose
probability distribution is like that of Eq.(\ref{ro}) with $\widehat{n}_{a}=%
\widehat{n}_{\vartheta }^{\downarrow }$. The unit vector $\widehat{n}%
_{\vartheta }^{\downarrow }$ is the one corresponding to the projector $%
\left| \downarrow \right\rangle _{\vartheta }{\vphantom{\left\langle
\downarrow \right|}}_{\vartheta }{\left\langle \downarrow \right| }$, with $%
\left| \downarrow \right\rangle _{\vartheta }=\cos \vartheta \left|
\downarrow \right\rangle _{z}-\sin \vartheta \left| \uparrow \right\rangle
_{z}=\delta \left| \downarrow \right\rangle _{z}-\gamma \left| \uparrow
\right\rangle _{z}$. Analogous prescriptions hold for exit port 4: change $%
\widehat{n}_{\vartheta }^{\downarrow }$ by $\widehat{n}_{\vartheta
}^{\uparrow }$, viz., change $\left| \downarrow \right\rangle _{\vartheta }$
by $\left| \uparrow \right\rangle _{\vartheta }$. In other words, the effect
of the Mach-Zehnder array is to flip the vector $\widehat{z}$ either to $%
\widehat{n}_{\vartheta }^{\uparrow }$ or to $\widehat{n}_{\vartheta
}^{\downarrow }$, depending on the exit channel. This is a perfectly
acceptable effect that such a device can have on variables like $\widehat{n}%
_{\lambda }$, irrespective of the physical meaning that we might
ascribe to these variables. As was shown in Ref. \cite{fdz}, the
KS\ model then predicts
that $\left\langle \widehat{\sigma }_{\theta }\right\rangle _{SG1}$ and $%
\left\langle \widehat{\sigma }_{\theta }\right\rangle _{SG2}$ will be given
by the corresponding expressions in Eqs.(\ref{sg1}) and (\ref{sg2}), thereby explaining the experimental outcomes as good as QM does.

In summary, an experiment that implements PH's proposal would produce
results that could be explained by a noncontextual realistic model, the KS\
model. Such an experiment would therefore be inconclusive about the
untenability of realistic, noncontextual models.

\end{document}